\documentclass{elsart}
\usepackage{amsmath}
\usepackage[ansinew]{inputenc}

\begin{document}
\begin{frontmatter}

\title{Optical studies of Cr$^{3+}$-Cr$^{2+}$ pair center in KZnF$_{3}$
crystal}

\author{M.V.Eremin, S.I.Nikitin, S.Yu.Prosvirnin,}
\author{N.I.Silkin, R.V.Yusupov}
\footnote{Corresponding author: Prof. M.V. Eremin, Kazan State
University, Kremlevskaya 18, 420008 Kazan, Russia. E-mail:
Mikhail.Eremin@ksu.ru; phone: (007 8432) 315116.}
\address{Kazan State University, Kremlevskaya 18, 420008 Kazan,
Russia}

\begin{abstract}
Optical absorption spectra of Cr$^{3+}$-Cr$^{2+}$ pair center in
KZnF$_{3}$:Cr$^{3+}$,Cr$^{2+}$ crystal were investigated in wide
temperature range. Broad band at 30800 cm$^{-1}$ is attributed to
cation-cation \(e_g\)-electron transfer transition. Narrow lines
with maxima at 16720 cm$^{-1}$ and 19880 cm$^{-1}$ have been
assigned to purely electronic exchange-induced electric-dipole
transitions from the ground
(Cr$^{3+}$,\(^4A_{2g}\);Cr$^{2+}$,\(^5E_g\)) state to excited
(Cr$^{3+}$,\(^4A_{2g}\);Cr$^{2+}$,\(^3E_g^a\)) and
(Cr$^{3+}$,\(^4A_{2g}\);Cr$^{2+}$,\(^3E_g^b\)) states,
respectively. It's vibronic satellites corresponding to \(a_{1g}\)
local mode of Cr$^{3+}$ fluorine octahedron of the pair are also
observed. Energy of the local mode for the ground and mentioned
excited states are 580, 540 and 530 cm$^{-1}$. Instead of expected
double exchange for mixed valence pair ferromagnetic superexchange
for Cr$^{3+}$-Cr$^{2+}$ pair in KZnF$_{3}$ crystal is realized.
Exchange integral \(J=-14.9\pm0.4\) cm$^{-1}$ and Jahn-Teller
splitting \(\Delta_{JT}=340\pm40\) cm$^{-1}$ for the ground state
of the pair were obtained by analysis of the temperature
dependence of absorption lines. Important features of the
crossover double exchange - ferromagnetic superexchange are
discussed.
\end{abstract}

\end{frontmatter}
\newpage
\section{Introduction}
Investigations of dynamic charge fluctuations in compounds like
La$_{2-x}$Sr$_{x}$CuO$_{4}$, NaV$_{2}$O$_{5}$,
La$_{1-x}$Ca$_{x}$MnO$_{3}$ are in the focus of condensed matter
physics. Physical essence of charge-spin dynamics in these systems
in our opinion can be revealed to some extent by investigations of
simple model mixed valence systems containing small number of
ions, the Hamiltonian of which may be solved exactly. For example
in the small polaron theory \cite{ref4,ref5} undoubtedly having
relationship to this problem main characteristics are introduced
in the two-site model.

In \cite{ref6,ref7,ref8} we have reported the observation of the
optical absorption lines of Cr$^{3+}$-Cr$^{2+}$ mixed valence pair
center in KZnF$_{3}$:Cr$^{3+}$,Cr$^{2+}$ crystals (Fig.1,a).
Optical piezospectroscopy investigations have shown that symmetry
of the pair is tetragonal. The most plausible model is that
chromium ions are located in neighboring sites along \(C_4\)-axis
of crystal. Observation of linear Stark effect on the absorption
lines of Cr$^{3+}$-Cr$^{2+}$ pair \cite{ref8} yields evidence that
the pair doesn't have center of symmetry at studied time scale.

In this paper we report the energy level structure of the ground
state of the pair. Several important characteristics as exchange
integral, energy of e$_{g}$-electron localization, frequencies of
local lattice vibrations near the pair center have been obtained.
We also point out important features of the crossover double
exchange - ferromagnetic superexchange in the ground state of
Cr$^{3+}$-Cr$^{2+}$ pair.
\section{Results and discussion}
Symmetry of KZnF$_{3}$ crystals is {\em Pm3m}, lattice constant is
{\em a}=4.055 \AA. The samples were grown by Bridgman-Stockbarger
method, for crystals doping CrF$_{3}$ and CrF$_{2}$ compounds were
used. The concentrations of chromium ions were varied in the range
0.1$\div$2 wt.\%. The concentrations of Cr$^{3+}$ and Cr$^{2+}$
ions in crystals were determined by measuring of absorption
coefficients of the bands corresponding to
$^{4}A_{2g}\rightarrow^{4}T_{2g}$ (14930 cm$^{-1}$) and
$^{5}E_{g}\rightarrow^{5}T_{2g}$ (11900 cm$^{-1}$) transitions of
single ions, respectively.

Absorption spectra were measured on the Specord-M40
spectrophotometer. For investigations of absorption spectra in
temperature range 4.2$\div$300 K Oxford Instruments CF-1204
optical cryostat was used.

Absorption spectrum of KZnF$_{3}$:Cr$^{3+}$,Cr$^{2+}$ crystal with
dopant's concentrations n(Cr$^{3+}$)= 1.8$\cdot$10$^{18}$
cm$^{-3}$ and n(Cr$^{2+}$)=9.2$\cdot$10$^{18}$ cm$^{-3}$ is
presented in Fig.2,a. On the wing of very intense band at $>$50000
cm$^{-1}$ a wide band at $\sim$43000 cm$^{-1}$ with the halfwidth
$\sim$7000 cm$^{-1}$ is observed. Studies of crystals containing
various concentrations of Cr$^{3+}$ and Cr$^{2+}$ ions have
displayed that intensities of these bands depend on concentration
of Cr$^{2+}$ ions only. Absorption bands observed at 59000
cm$^{-1}$ and 46300 cm$^{-1}$ in the spectrum of
KMgF$_{3}$:Cr$^{2+}$ crystal were assigned to transitions from
\(3d\) to \(4p\) and \(4s\) states of Cr$^{2+}$ ions, respectively
\cite{ref9,ref10}. Since KMgF$_{3}$ crystal is isostructural to
KZnF$_{3}$ it is natural to conclude that absorption bands of
KZnF$_{3}$:Cr$^{3+}$,Cr$^{2+}$ crystal at $>$50000 cm$^{-1}$ and
$\sim$43000 cm$^{-1}$ should be attributed to \(3d\rightarrow4p\)
and \(3d\rightarrow4s\) transitions of single Cr$^{2+}$ ions,
respectively.

In crystals with bigger concentrations of Cr$^{3+}$ ions
(absorption spectrum of a sample with
n(Cr$^{3+}$)=1.3$\cdot$10$^{20}$ cm$^{-3}$ and
n(Cr$^{2+}$)=7.3$\cdot$10$^{18}$ cm$^{-3}$ is shown in Fig.2,b)
additional absorption band at 30800 cm$^{-1}$ with the halfwidth
$\sim$7000 cm$^{-1}$ (except of well known \(d-d\) transitions of
single Cr$^{3+}$ ions \cite{ref16}) has been observed. Intensity
of this band correlates to intensities of absorption lines of
Cr$^{3+}$-Cr$^{2+}$ pair center in the visible (Fig.1,a). It's
oscillator strength to our estimates is \(f\sim10^{-1}\). These
facts allowed us to assign absorption band with
\(\nu_{max}\sim30800\) cm$^{-1}$ to charge-transfer transition
(Cr$^{2+}$, \(t_{2g}^{3},e_{g}:\) $^{5}$E$_{g}$; Cr$^{3+}$,
\(t_{2g}^{3}:\) $^{4}$A$_{2g}$)$\rightarrow$ (Cr$^{3+}$,
\(t_{2g}^{3}:\) $^{4}$A$_{2g}$; Cr$^{2+}$, \(t_{2g}^{3},e_{g}:\)
$^{5}$E$_{g}$) of Cr$^{3+}$-Cr$^{2+}$ pair center.

High energy of this transition reveals strong interaction of the
pair center with the lattice. Indeed, let us consider the small
polaron two-site model \cite{ref4,ref5} taking into account the
appropriate electronic states of Cr$^{3+}$ and Cr$^{2+}$ ions.
Hamiltonian of the system in the second quantization form is
written as:
\begin{equation}\label{Hamilton}
  \hat{H}=\frac{p^{2}}{M}+\frac{M\omega^{2}}{2}q^{2}+
  \frac{Vq}{2}(a^{+}a-b^{+}b)+t(a^{+}b+b^{+}a),
\end{equation}
where {\em q} is a vibrational coordinate describing the
difference in local surrounding displacements corresponding to the
localization of "extra" {\em e$_{g}$}-electron at a or b Cr$^{3+}$
ion:
\begin{equation}\label{Qs}
  q=Q_{a}-Q_{b},
\end{equation}
{\em p} - momentum canonically conjugated to coordinate {\em q};
{\em V} is a parameter of linear vibronic interaction with {\em
Q$_{a}$} and {\em Q$_{b}$} modes of fluorine octahedra of a and b
ions; {\em a, a$^{+}$, b, b$^{+}$} - creation and annihilation
operators of {\em e$_{g}$}-electron at ions a and b, respectively;
{\em t} is a transfer integral.

In the case {\em q=0}, Hamiltonian (\ref{Hamilton}) corresponds to
usual double exchange, it's diagonalization was described in
\cite{ref11}. Energy level scheme of the system depends on total
spin of the pair and is equidistant:
\begin{equation}\label{DE}
  \varepsilon(S)=\pm\tilde{t}(S+\frac{1}{2}),
\end{equation}
where \(\tilde{t}=t/(2S_0+1)\) , {\em S$_{0}$} is ion core spin.
In our case the ion core is \((t_{2g}^3:^4A_{2g})\) and {\em
S$_{0}$}=3/2.

For {\em q$\neq$0}, diagonalization of (\ref{Hamilton}) gives the
following expression for adiabatic potentials:
\begin{equation}\label{Adiab}
  E^{\pm}(q,S_{0})=\frac{M\omega^{2}q^{2}}{4}\pm
  \sqrt{\frac{V^{2}q^{2}}{4}+\varepsilon^{2}(S_{0})}.
\end{equation}
As one can see, when \(\varepsilon(S)\gg Vq\) the adiabatic
potential has a minimum at \(q=0\). This case corresponds to
modified double exchange, the energy spectrum is still
equidistant. If the vibronic energy is larger than
\(\varepsilon(S)\) the adiabatic potential has two wells as it is
shown in Fig.3 (vertical arrow indicates possible charge-transfer
transition). This case corresponds to partial localization of
"extra" \(e_g\)-electron at a or b centers. For adiabatic
potential in Fig.3 the following set of parameters was used: the
transfer integral between \((3z^2-r^2)\) orbitals via intermediate
fluorine ion \(t_{uu}=2400\) cm$^{-1}$ as it is found in
\cite{ref12} from analysis of the optical spectra of
exchange-coupled Cu$^{2+}$-Mn$^{2+}$ pairs in KZnF$_{3}$ crystal,
\(V\) was obtained from the energy of charge-transfer transition
\(Vq_0\)=30800 cm$^{-1}$, where \(q_0\) is equilibrium coordinate,
\(\hbar\omega=580\) cm$^{-1}$ and will be explained below.

In the minima of adiabatic potentials \(e_g\)-electron is mainly
localized at one of the ions of the pair. Energy spectrum
corresponding to minima of adiabatic potentials obeys Lande
intervals rule
\begin{equation}\label{Lande}
    E^{-}(q_{0},S)-E^{-}(q_{0},S-1)\approx-\frac{2\tilde{t}^{2}}
    {Vq_{0}}S,
\end{equation}
and hence can be described as superexchange.

We can say that electron-lattice interaction suppresses double
exchange modifying it to ferromagnetic superexchange. As it can be
easily seen from (\ref{Adiab}), crossover double exchange -
ferromagnetic superexchange happens rather sharply with vibronic
coupling increase. These features of crossover haven't been
mentioned in literature before but are very important in our
opinion.

The intensities of absorption lines 1 (\(\nu_{max}\)=16720
cm$^{-1}$, Fig.1), 2 (19880), 1$^{\prime}$ (17260), 2$^{\prime}$
(20410), 1$^{\prime\prime}$ (16140) and 2$^{\prime\prime}$ (19300)
are proportional to the product of Cr$^{3+}$ and Cr$^{2+}$ ions
concentrations and therefore these lines were assigned to
Cr$^{3+}$-Cr$^{2+}$ pair center \cite{ref6}. The estimate of
oscillator strengths of lines 1 and 2 as \(f\sim5\cdot10^{-4}\)
shows that these lines can be attributed to electric-dipole
transitions. Analysis of exchange-induced electric-dipole
transitions of Cr$^{3+}$-Cr$^{2+}$ pair center\footnote{Detailed
analysis of transitions exceeds the limits of present work and
will be reported elsewhere \cite{ref15}} including relative
intensities and selection rules according to \cite{ref13} permits
us to unambiguously assign lines 1 and 2 to purely electronic
transitions from the ground (Cr$^{3+}$:$^{4}$A$_{2g}$;
Cr$^{2+}$:$^{5}$E$_{g}$) to excited (Cr$^{3+}$:$^{4}$A$_{2g}$;
Cr$^{2+}$:$^{3}$E$_g^a$) and (Cr$^{3+}$:$^{4}$A$_{2g}$;
Cr$^{2+}$:$^{3}$E$_g^b$) states, respectively. These excited
states have nearly the same electronic configurations
(Cr$^{3+}$,\(t_{2g}^{3}\); Cr$^{2+}$,\(t_{2g}^{3}e_g\)) as the
ground state, total spin is changed only. As a consequence the
adiabatic potentials of the ground and given excited states are
almost identical. In this way we explain why the absorption lines
1 and 2 are narrow.

Lines 1$^{\prime}$ and 2$^{\prime}$ are observed in the spectrum
equal energy intervals apart from lines 1 and 2. Their temperature
dependencies are within a constant factor the same as those for
lines 1 and 2 (Fig.4). To the lower energies from absorption lines
1 and 2 lines 1$^{\prime\prime}$ and 2$^{\prime\prime}$ clearly
displayed in the derivative of absorption spectrum (Fig.1, b) at
T\(=300\) K are observed. These lines disappear when temperature
is lowered to T\(\sim150\) K (Fig.1, c) and therefore can be
assigned to transitions including excited vibronic states. Lines
1$^{\prime}$ and 2$^{\prime}$ correspond to transitions with
\(\Delta{n}=1\), where \(n\) is vibrational quantum number, lines
1$^{\prime\prime}$ and 2$^{\prime\prime}$ - to transitions with
\(\Delta{n}=-1\). Energies of vibronic modes for the ground and
the excited states can be easily determined from spectrum and are
equal to \(\hbar\omega_g\)=580$\pm$20 cm$^{-1}$,
\(\hbar\omega_1\)=540$\pm$20 cm$^{-1}$ and
\(\hbar\omega_2\)=530$\pm$20 cm$^{-1}$, respectively. It is
interesting to point that \(\hbar\omega_g\) is equal to the energy
of local vibration mode of \(a_{1g}\) symmetry as it is reported
for single Cr$^{3+}$ ion in KZnF$_{3}$ crystal (574 cm$^{-1}$,
\cite{ref14}). This fact allows us to conclude that \(Q_a\) and
\(Q_b\) in (\ref{Qs}) are mainly "breathing" modes of the fluorine
octahedra near the ions of the pair. It should be mentioned that
condensation of local vibration due to localization of extra
charge on the lattice site was expected and discussed in theory
(see for example \cite{ref5}), but to our knowledge we report here
their so clear experimental observation for the first time.

Intensities of absorption lines 1, 2, 1$^{\prime}$ and
2$^{\prime}$ depend strongly on temperature. They are not
observable at T$<$15 K. When temperature increases intensities of
these lines grow, show up a maximum at T$\sim$150 K and decrease
slowly with further temperature increase (Fig.4). Strong
temperature dependence of absorption lines of Cr$^{3+}$-Cr$^{2+}$
pair center in the temperature range T$<$120 K can be naturally
explained by ferromagnetic type of exchange interaction in the
ground state. In this case total spin state with S=\(\frac{7}{2}\)
is the ground state. Electric-dipole transitions from this
sublevel obeying the selection rule \(\Delta{S}=0\) are forbidden
because in excited states
(Cr$^{3+}$:$^{4}$A$_{2g}$;Cr$^{2+}$:$^{3}$E$_{g}^{a,b}$) of the
pair states with S=\(\frac{7}{2}\) are absent. Absorption appears
due to thermal occupation of the excited states with
S=\(\frac{5}{2}\), \(\frac{3}{2}\) and \(\frac{1}{2}\).

The quantitative analysis of temperature dependence of absorption
lines was successfully performed under following conclusions. The
localization energy (or polaronic shift) \(Vq_0/4\sim7700\)
cm$^{-1}$ is large enough with respect to \(e_g\)-electron
transfer energy and as a result we have the case of ferromagnetic
superexchange instead of expected double exchange. This conclusion
is supported by the observation in absorption spectrum of vibronic
satellites. Transitions from excited vibronic sublevel were also
taken into account. Parameters \(Vq_0\) and \(t_{uu}\) for this
state were taken the same as for the ground vibronic state. The
splittings of the ground state spin multiplets \(S=\frac{7}{2}\),
\(S=\frac{5}{2}\) and \(S=\frac{3}{2}\) due to spin-orbit
interaction and axial component of crystal field are small in
respect to exchange splittings \cite{ref15} and were neglected.
Relative probabilities of transitions from different total spin
states were found in \cite{ref6}:
\begin{equation}
  W(\frac{5}{2}):W(\frac{3}{2}):W(\frac{1}{2})=21:16:5.
\end{equation}
Result of the fit within the proposed model is shown in Fig.4 by
dashed line. For the temperature range T$<$120 K observed
temperature dependence is fairly well fitted, but at higher
temperatures significant disagreement appears. This disagreement
hints at the importance of Jahn-Teller nature of [CrF$_6$]$^{4-}$
fragment of the pair center, which was not treated here up to now.
Due to the presence of Cr$^{3+}$ ion in the pair center three
minima of adiabatic potential of [CrF$_6$]$^{4-}$ cluster have
different energies. Minimum corresponding to elongation of the
octahedron towards Cr$^{3+}$ ion is the deepest, \(e_g\)-electron
resides at \((3z^2-r^2)\) orbital. Other two minima which
correspond to perpendicular elongations of fluorine octahedron are
equivalent and a little bit higher in energy (energy gap between
the minima is \(\Delta_{JT}\)). It is obvious that the transfer
integral between \((3x^2-r^2)\) and \((3y^2-r^2)\) orbitals of
chromium ions due to their small overlapping with \(p\)-orbitals
of intermediate F$^-$ ion will be much smaller than that for
\((3z^2-r^2)\) orbital. For this reason exchange splitting of
these Jahn-Teller states will be relatively small. Since
\(e_g-e_g\) kinetic exchange is the main mechanism of the
exchange-induced optical absorption, it is clear that described
Jahn-Teller excited states do not contribute to absorption
spectrum. The excited state at energy \(\Delta_{JT}\) is
forty-fold degenerate. It's thermal occupation yields to the drop
of absorption lines intensities at T$>$150 K.

Final fit of the temperature dependence including Jahn-Teller
states as considered above is shown in Fig.4 by solid line. The
values of extracted parameters are: \(J=-14.9\pm0.4\) cm$^{-1}$,
\(\Delta_{JT}=\)340$\pm$40 cm$^{-1}$. As one can see, fitting
curve fairly well follows experimental dependence in the whole
temperature range.
\section{Summary}
Main results of the paper can be summarized as follows. Absorption
band with \(\nu_{max}\sim30800\) cm$^{-1}$ in
KZnF$_3$:Cr$^{3+}$,Cr$^{2+}$ crystal spectrum corresponds to
cation-cation charge-transfer transition from Cr$^{2+}$ ion to
Cr$^{3+}$. Large charge-transfer energy reveals high degree of
extra charge localization at one of the ions of the pair caused by
strong vibronic interaction. Energy of charge localization (or
polaronic shift) is 7700 cm$^{-1}$. Absorption lines of
Cr$^{3+}$-Cr$^{2+}$ pair at 16720 cm$^{-1}$ and 19880 cm$^{-1}$
are associated with Cr$^{2+}$ ion excitation from the ground
$^5$E$_g$ to excited $^3$E$_g^a$ and $^3$E$_g^b$ states. Energies
of local vibrations in the ground and excited states of the pair
are found to be 580$\pm$20 cm$^{-1}$, 540$\pm$20 cm$^{-1}$,
530$\pm$20 cm$^{-1}$, respectively.

Temperature dependence of integral intensities of absorption lines
has been investigated. The evidence of Jahn-Teller effect in
Cr$^{3+}$-Cr$^{2+}$ pair has been found; energy gap between
Jahn-Teller minima of adiabatic potential of the pair center has
been estimated as \(\Delta_{JT}=340\pm40\) cm$^{-1}$.

Important features of transitive region from double exchange to
superexchange are pointed out. In particular it has been
demonstrated that strong vibronic interaction in mixed valence
pair center leads to crossover double exchange - ferromagnetic
superexchange. Exchange integral for the ground state is
\(J=-14.9\pm0.4\) cm$^{-1}$.

The work was supported by Russian Foundation for Basic Researches,
grant No.00-02-17597.

\newpage
\pagestyle{empty}
\title{Figure captions}

Fig.1. Absorption spectrum of KZnF$_3$:Cr$^{3+}$,Cr$^{2+}$ crystal
at \(T=300\) K (a); derivatives of absorption spectra at \(T=300\)
K (b) and \(T=150\) K (c).

Fig.2. Absorption spectra of KZnF$_3$:Cr$^{3+}$,Cr$^{2+}$ crystals
with chromium ions concentrations n(Cr$^{3+}$)=\(1.8\cdot10^{18}\)
cm$^{-3}$, n(Cr$^{2+}$)=\(9.2\cdot10^{18}\) cm$^{-3}$ (a,1) and
n(Cr$^{3+}$)=\(1.3\cdot10^{20}\) cm$^{-3}$,
n(Cr$^{2+}$)=\(7.3\cdot10^{18}\) cm$^{-3}$ (b,1); \(T=300\) K.
Curves a,2 and b,2 are results of subtraction of the wing of
absorption band with \(\nu_{max}>50000\) cm$^{-1}$ (dashed line)
and \(d-d\) bands of Cr$^{3+}$ ions.

Fig.3. Adiabatic potential for Cr$^{3+}$-Cr$^{2+}$ pair center in
KZnF$_3$ crystal; charge-transfer transition is shown by the
arrow.

Fig.4. Temperature dependence of integral intensity of
Cr$^{3+}$-Cr$^{2+}$ pair absorption line at 16720 cm$^{-1}$ and
its fitting by the models of the ground state with (a) and without
(b) taking into account Jahn-Teller interaction.

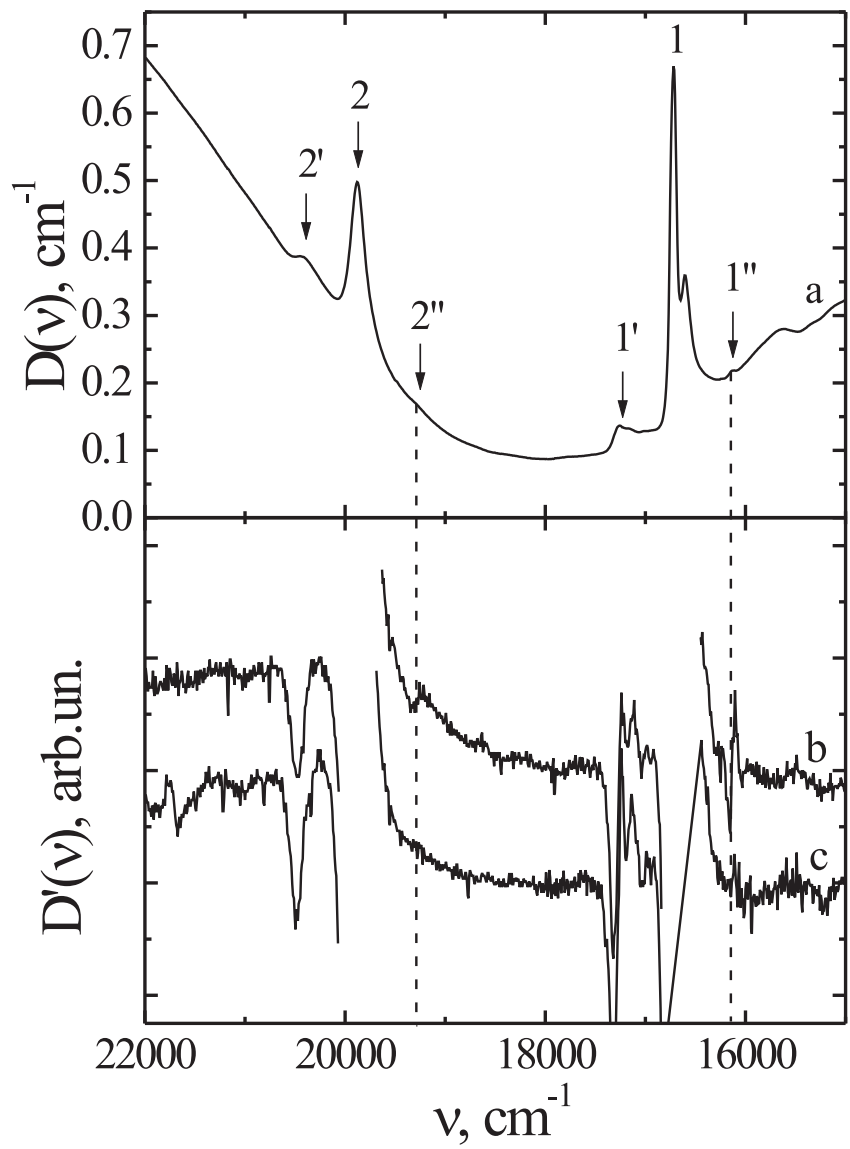
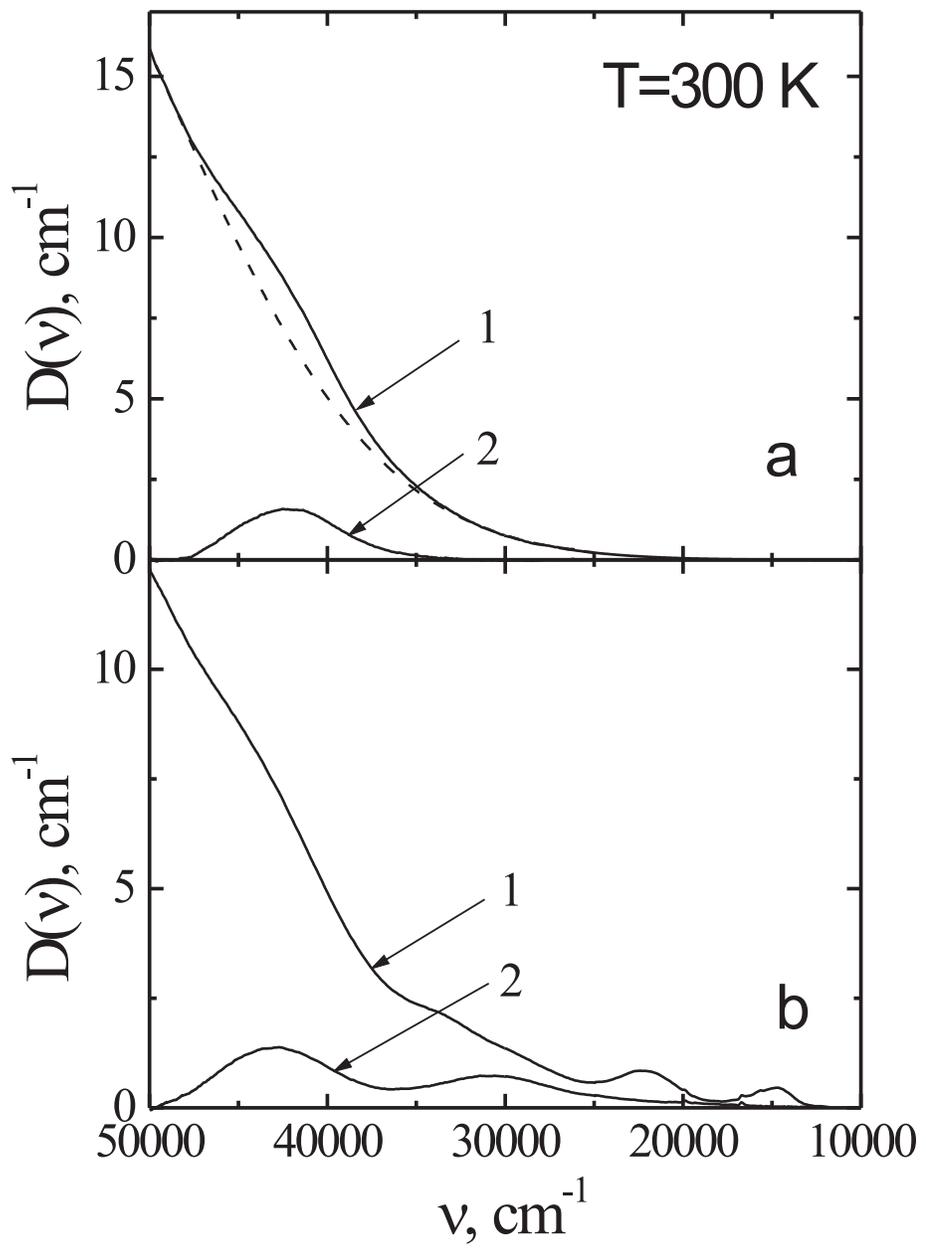
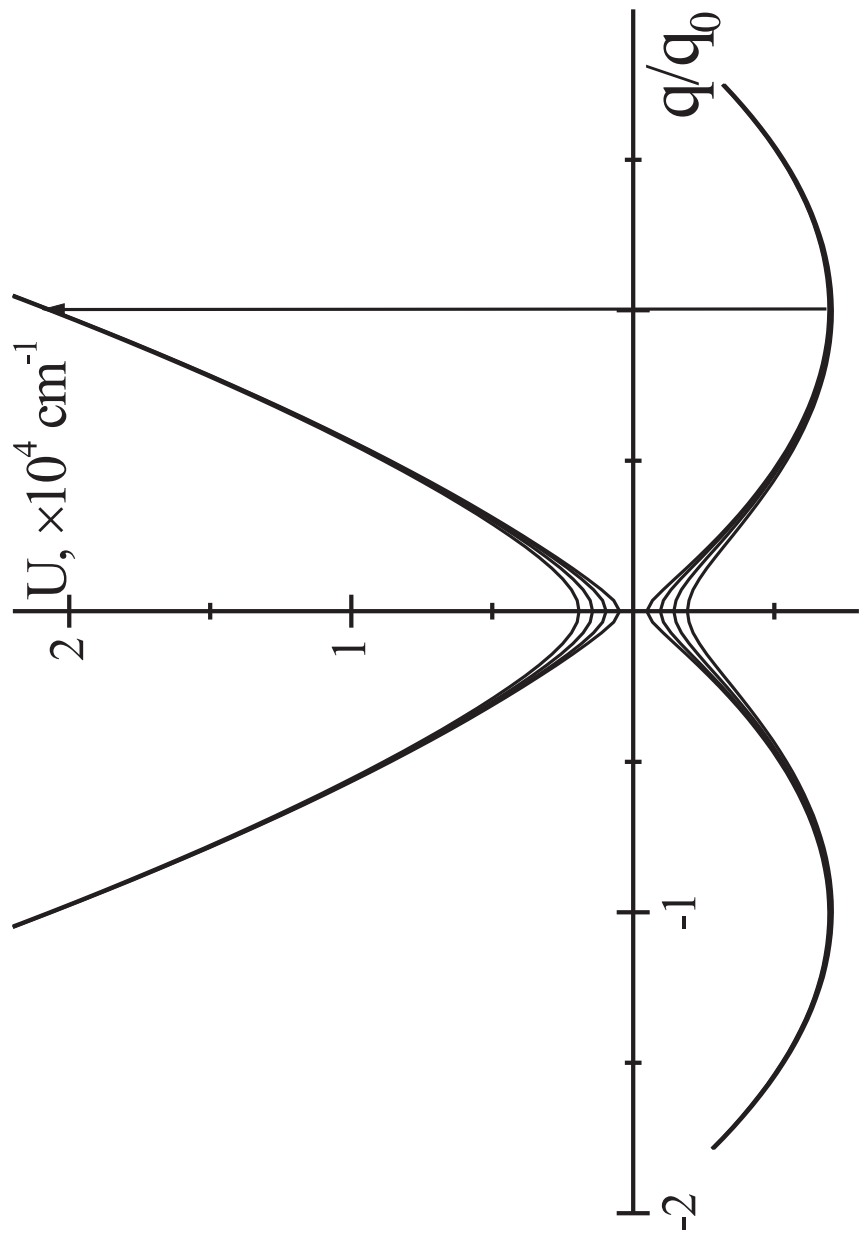
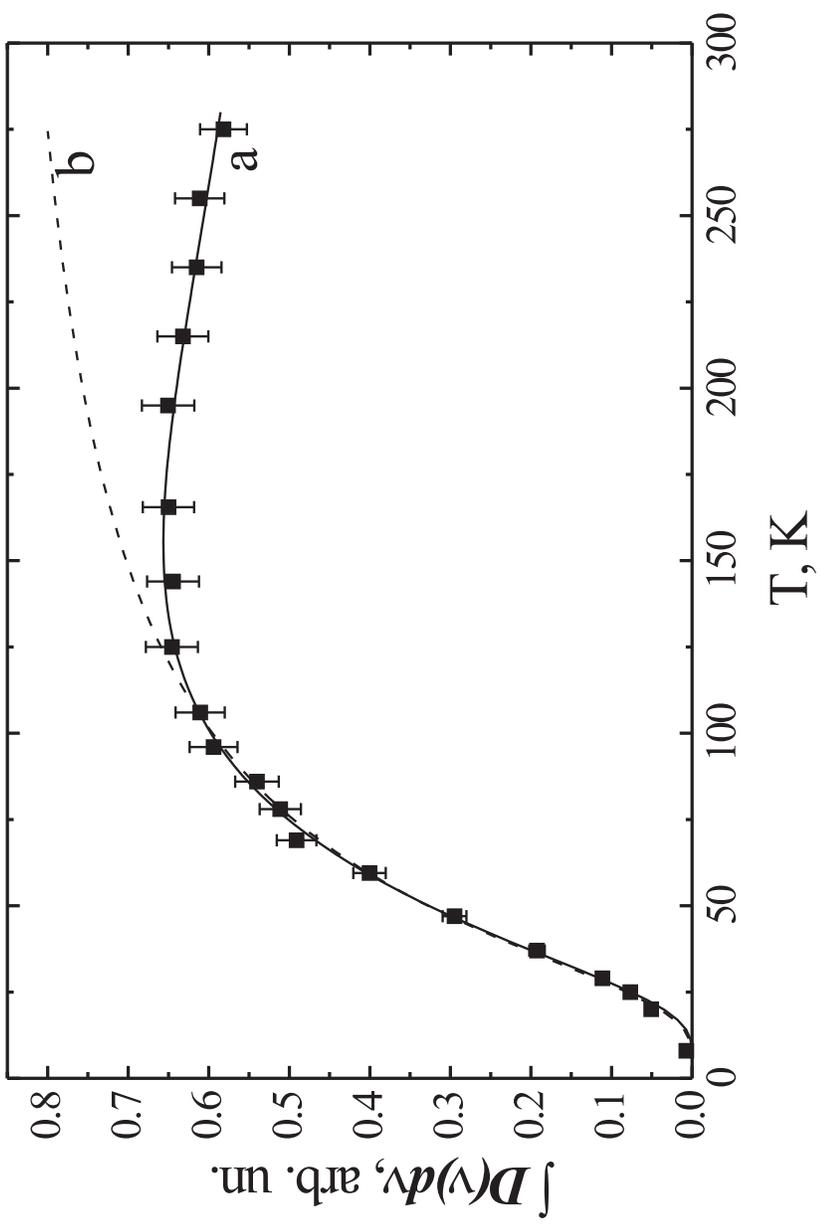
\end{document}